\documentclass[a4paper]{jpconf}
\usepackage{amssymb,amsmath}
\usepackage{color}
\usepackage{accents}
\usepackage{texdraw}
\usepackage{eucal}
\usepackage{epic,epsfig}
\usepackage{graphicx}
\usepackage{floatrow}

\newcommand{\ut}{\undertilde}

\begin{document}
\title{Ground state entanglement entropy for discrete-time two coupled harmonic oscillators}

\author{Watcharanon Kantayasakun$^{\dagger,1 } $, Sikarin Yoo-Kong$^{\dagger,\ddagger,\amalg,2} $, Tanapat Deesuwan$^{\dagger,\odot,3} $, Monsit Tanasittikosol$^{\dagger,4} $,Watchara Liewrian$^{\dagger,\ddagger,5} $}

\address{$^\dagger$ Theoretical and Computational Physics (TCP) Group, Department of Physics, Faculty of Science, King Mongkut's University of Technology Thonburi, Bangkok 10140, Thailand. \\
$^\ddagger$ Theoretical and Computational Science Centre (TaCs), Faculty of Science, King Mongkut's University of Technology Thonburi, Bangkok 10140, Thailand.\\
$^\amalg$ Ratchaburi Campus, King Mongkut's University of Technology Thonburi, Ratchaburi, 70510, Thailand.\\
$^\odot$  Learning Institute, King Mongkut's University of Technology Thonburi, Bangkok 10140, Thailand.}

\ead{$^1$moontown\_lop@hotmail.com,
\small  $^2$syookong@gmail.com,
\small  $^3$tanapat.deesuwan@gmail.com,
\small  $^4$monsit.tan@gmail.com,
\small  $^5$watchara.lie@kmutt.ac.th}

\begin{abstract}
The ground state entanglement of the system, both in discrete-time and continuous-time cases, is quantified through the linear entropy. The result shows that the entanglement increases as the interaction between the particles increases in both time scales. It is also found that the strength of the harmonic potential affects the formation rate of the entanglement of the system. The different feature of the entanglement between continuous-time and discrete-time scales is that, for discrete-time entanglement, there is a cut-off condition. This condition implies that the system can never be in a  maximally entangled state.       
\end{abstract}

\section{Introduction}
The idea that time flow constitutes from discrete-steps was suggested by many physicists \cite{AE,RP,TD,GH}. The question which could be raised from this idea is whether or not there are similarities or differences of the physical behaviors at the discrete-time scale and continuous-time scale. To answer this question, the system of two coupled harmonic oscillators is used as a toy model to study at the quantum level. The comparison between the discrete-time wave function and the continuous-time wave function is investigated. Furthermore, an important feature in quantum mechanics called entanglement is examined in detail. What we expect to observe in this study are some extra-features arising due to the discreteness of the time flow.

The organisation of this article is as follows. In Section 2, the formulation of the equations of motion of the two coupled oscillators is set up in both discrete-time and continuous-time scales. Then, in Section 3, the discrete-time wave function is computed and with the modified uncertainty principle. Once the wave function is obtained, the linear entanglement entropy is computed in Section 4 together with the discussion. Finally, the conclusion is provided with some remarks.

\section{Discrete-time coupled harmonic oscillators}
The system consists of two identical particles with unit mass and the interactions between themselves, and, between the particles and the environment are modelled by Hooke's law with coupling constants $\sigma$ and $k$, respectively. The Hamiltonian of the system is given by
\begin{equation}\label{H1}
H(p_1,p_2,x_2,x_2)=p^2 _1/2+p^2 _2/2+kx^2_1/2+kx^2 _2/2+\sigma(x_1-x_2)^2/2\;,
\end{equation}
where $p_{ i}$ and $x_{i}$ are the momentum and position of the $i^{\rm th}$ particle and $i=1,2$. To decouple the Hamiltonian, the normal coordinates $X_{1}=(x_{1}+x_{2})/\sqrt{2}$ (mode 1) and $X_{2}=(x_{1}-x_{2})/\sqrt{2}$ (mode 2) are used to transform Eq. (\ref{H1}) into
\begin{equation}\label{H2}
H(P_1,P_2,X_1,X_2)=\left(P_{1}^2 +\omega^2_{1}X_{1}^2 \right)/2+\left(P_{2}^2 +\omega^2_{2}X_{2}^2 \right)/2\;,
\end{equation} 
where $P_{ i}$ are new momentum variables and the angular frequencies are  $\omega_{1}=\sqrt{k}$ (mode 1) and $\omega_{2}=\sqrt{(k/2+\sigma)2}$ (mode 2). We now introduce the discrete-time Hamiltonian \cite{CM,MV} given by
\begin{equation}\label{H3}
H(\tilde{P}_{1},\tilde{P}_{2},X_{1},X_{2})=\left(\tilde P_{1}^2 +\omega^2_{1}X_{1}^2 \right)/2+\left(\tilde P_{2}^2 +\omega^2_{2}X_{2}^2 \right)/2\;,
\end{equation}
where $P_{ i}(n)$ and $X_{ i}(n)$ are the discrete-time momentum and position of the ${ i}^{\rm th}$ particle at time $n$. The shifted momentum is $\tilde P_i=P_i(n+\epsilon)$, where $\epsilon$ is the discrete-time step. The discrete-time Hamilton equations are $\partial{H}/\partial{\tilde{P}_{i}}=( \tilde{X}_{i}-{X}_{i}) / \epsilon$ and $\partial{H}/ \partial{X}_{i}=-( \tilde{P}_{i}-{P}_{i}) / \epsilon$, resulting in discrete maps
\begin{equation}\label{DM}
\tilde{X}_{i}=(1-\omega^2_{i}\epsilon^2)X_{i}+P_{i}\epsilon\;,\;\;\mbox{and}\;\;
\tilde{P}_{i}=-\omega^2_{i}\epsilon X_{i}+P_{i}\;.
\end{equation}
Eliminating the momentum variable $P_{i}$, we obtain the discrete-time equation of motion $\tilde{X}_{i}+\ut{X}_{i}=2(1-\omega^2_{i}\epsilon^2/2)X_{i}$ of the system. Note that, under the continuum limit $\epsilon\to0$, the continuous-time equation of motion for the system is recovered. 

\section{Discrete-time wave function}
To obtain the discrete-time wave function, we start with the function \cite{CMF}
\begin{equation}\label{In}
\hat{I}_i=\hat{P}_{i}^2+\omega^2_{i}\hat{X_{i}}^2-\epsilon\omega^2 _{i} \left(\hat{P_{i}}\hat{X_{i}}+\hat{X_{i}}\hat{P_{i}} \right)/2\;,
\end{equation} 
where $\hat P_{i}$ and $\hat X_{i}$ are operators. Since $\hat {I}_i$ is invariant under the map (\ref{DM}): $ \tilde{\hat{I}}_i=\hat{I}_i$, then it can be treated as the effective Hamiltonian resulting in    
\begin{equation}\label{Sch}
\hat{I}\Psi(X_1,X_2) = \left[\sum_{i=1}^{2} \left( -\hbar^2\frac{\partial^2}{\partial X_{i}^2}+{\rm i}\epsilon\omega^2_{i}\hbar X_{i}\frac{\partial }{\partial X_{i}}+\omega^2_{i}X_{i}^2+\frac{{\rm i}\hbar}{2}\epsilon\omega^2_{i}\right) \right]\Psi(X_1,X_2)=E\Psi(X_1,X_2)\;,
\end{equation}
where $\hat{I}=\hat{I}_{1}+\hat{I}_{2}$ is the total effective Hamiltonian operator and $E$ is the total energy of the system. Writing the wave function as $\Psi(X_1,X_2)=\psi(X_{1})\varphi(X_{2})$ and using the transformations $\varphi(X_{1})=w(X_{1}){\rm exp}\left[\rm i\epsilon\omega^2_{1}X_{1}^2/(4\hbar)\right]$ and $\varphi(X_{2})=w(X_{2}){\rm exp}\left[\rm i\epsilon\omega^2_{2}X_{2}^2/(4\hbar)\right]$, the eigenfunctions of $\hat{I}$ are
\begin{equation}\label{WF}
\Psi_{nm}=\left(\frac{\Omega_{1}}{\pi\hbar}\right) ^{\frac{1}{4}}\left(\frac{\Omega_{2}}{\pi\hbar}\right) ^{\frac{1}{4}}\frac{1}{\sqrt{2^n n!}}\frac{1}{\sqrt{2^m m!}}H_{n}\left(\sqrt{\frac{\Omega_{1}}{\hbar}}X_{1}\right)  H_{m}\left(\sqrt{\frac{\Omega_{2}}{\hbar}}X_{2}\right) {\rm  e}^{(\frac{{\rm i}\epsilon\omega^2_{1}}{4\hbar}-\frac{\Omega_{1}}{2\hbar})X_{1}^2+(\frac{{\rm i}\epsilon\omega^2_{2}}{4\hbar}-\frac{\Omega_{2}}{2\hbar})X_{2}^2}
\end{equation}
where $n,m=0,1,2,3,....\;$ and $H_y(x)$ is the Hermite Polynomial of order y.
\begin{figure}[h]
\begin{center}
\includegraphics[scale=0.3]{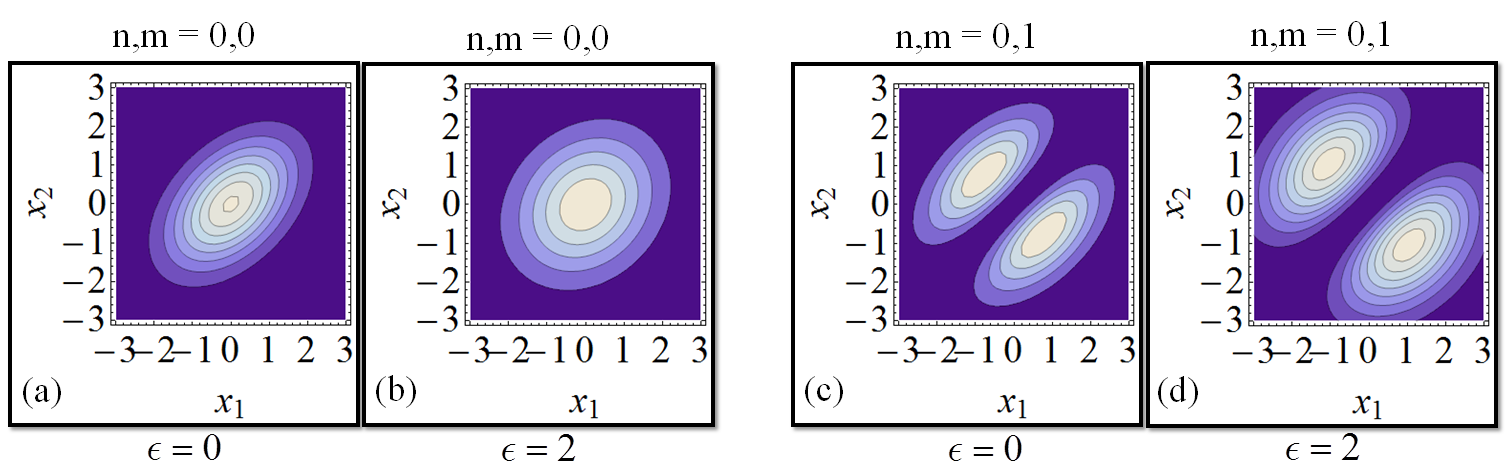}
\caption{\label{CT} Contour plots of the probability density for ground state (a),(b) and first excited state (c),(d) for $k=0.1$ and $\sigma=0.3$.}
\end{center}
\end{figure}
The total energy now is $E_{nm}=E_{n}+E_{m}$, where $E_{n}=2\hbar\Omega_{1}\left( n+1/2\right)$ and $\Omega_{1}=\omega_{1}\sqrt{\left( 1-\epsilon^2 \omega^2 _{1}/4\right)}$ (mode 1), and, $E_{m}=2\hbar\Omega_{2}\left( m+1/2\right)$ and $\Omega_{2}=\omega_{2}\sqrt{\left( 1-\epsilon^2 \omega^2 _{2}/4\right)}$ (mode 2). Under the continuum limit $\varepsilon\to0$, the wave function (\ref{WF}) is identical to that of the continuous-time harmonic oscillators.
The contour of the probability density is shown in Fig. \ref{CT} for the case of $\epsilon=0$ (continuous-time case), and $\epsilon=2$. According to Fig. 1, the probability of the discrete-time wave function is a little bit broader than that of the continuous-time wave function. This results from the fact that both the exponential terms ${\rm exp}\left[ -\Omega_{i}X^{2}_{i}/(2\hbar)\right]$ and the Hermite Polynomials $H_{y}(\sqrt{\Omega_{i}/\hbar}X_{i})$ contain the discrete-time parameter $\epsilon$. Furthermore, we find that the uncertainty principle for each mode in this discrete-time setting is altered to
\begin{equation}\label{UN}
\sigma_{X_{i}}\sigma_{P_{i}}=\hbar\left(y_i+1/2 \right)\sqrt{1+{\epsilon^2 \omega^4 _{i}}/{4\Omega^2 _{i}}}\;,
\end{equation}
where $y_1=n$ and $y_2=m$. This leads to the modified Heisenberg algebra $\left[ X_{i},P_{i}\right]\;=i\hbar\sqrt{1+{\epsilon^2 \omega^4 _{i}}/{4\Omega^2 _{i}}}\; $.

\section{Entanglement entropy of the ground state}
To study the entanglement behavior of the ground state of quantum discrete-time coupled harmonic oscillators, we use the linear-entropy $S_{L}$ given by
\begin{equation}
S_{L_{j}}=1-\Tr(\rho^2_{j})\;,
\end{equation}
where $j=1,2$, $\rho_{j}=\Tr_{\frac{2}{j}}\rho_{12}=\int \rho_{12} dx_{\frac{2}{j}}$ is the reduced density matrix of the system $j$, and $\rho_{12}$ is the full density matrix.  Note that, for a global pure state, the linear-entropy of the reduced state is bounded between $0\leq S_{L} \leq1$, where $S_L=1$ indicates the whole system is  maximally entangled and $S_L=0$ indicates a separable state. The full density matrix of the ground state is $\rho_{12}(x_{1},x_{2};x'_{1},x'_{2})=\Psi_{00}(x_{1},x_{2})\Psi^\ast_{00}(x'_{1},x'_{2})$ and therefore
\begin{equation}
S_{L}=1-\frac{\gamma-\beta}{\sqrt{\gamma^2 -\beta^2}} \;,
\end{equation}
where
\begin{equation}
\gamma=\frac{1}{4}(\Omega_{1}+\Omega_{2})+\frac{\Omega_{1}\Omega_{2}}{\Omega_{1}+\Omega_{2}}+\frac{\epsilon^2 \sigma^2}{4(\Omega_{1}+\Omega_{2})}, \beta=\frac{1}{4}(\Omega_{1}+\Omega_{2})-\frac{\Omega_{1}\Omega_{2}}{\Omega_{1}+\Omega_{2}}+\frac{\epsilon^2 \sigma^2}{4(\Omega_{1}+\Omega_{2})}.
\end{equation}
\begin{figure}[h]
\begin{center}
\includegraphics[scale=0.4]{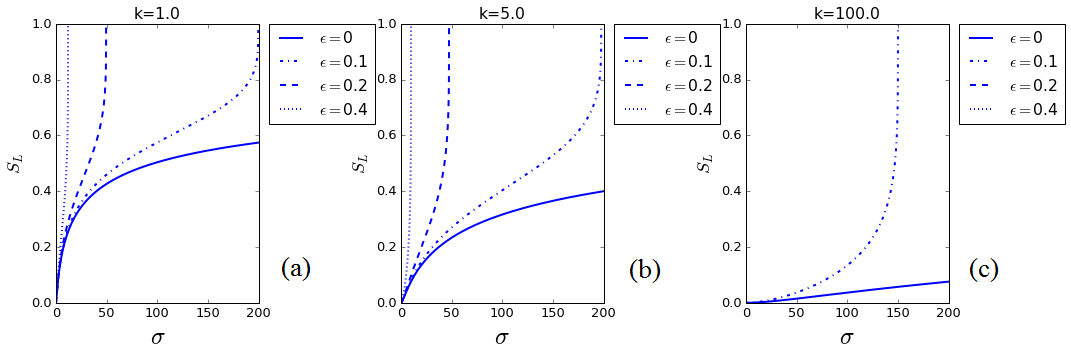}
\caption{\label{ES} The relation between the linear entropy and the internal interaction ($\sigma$) with different amount of the discrete-time scale and the external interaction ($k$).}
\end{center}
\end{figure}
\\
According to Fig. \ref{ES}, in the continuous time(solid lines), the entanglement of the system at the ground state increases as the interaction between particles $\sigma$ increases, while the interaction with environment $k$ is fixed. The system approaches to the maximally entangled state $S_L\rightarrow 1$ as the parameter $\sigma$ approaches to infinity implying that the oscillation mode $\Omega_1$ (the center of mass motion) significantly dominates over the oscillation mode $\Omega_2$ (the relative motion). We also find that when the parameter $k$ increases, the entanglement will rise more slowly with the increasing value of the parameter $\sigma$. This means that the oscillation mode $\Omega_2$ becomes more significant with increasing $k$ which then makes the oscillation mode $\Omega_1$ more difficult to overcome the oscillation mode $\Omega_2$. In the case that the parameter $k$ is infinitely large, the entanglement of the system is extremely suppressed due to the domination of the oscillation mode $\Omega_2$. We may now say that less relative motion (the oscillation mode 2) of the system implies more entanglement.

In the discrete time case, the entanglement of the system behaves almost the same with the continuous time case. Except that we cannot freely vary the values of the parameter $\sigma$ and the parameter $k$ since there are the cut-off conditions coming from the fact that both $\Omega_{1}$ and $\Omega_{2}$ must be positive values. This implies that     
$0\leq\omega_2^2<4/\epsilon^2$ since $\omega_{2}\geq \omega_{1}$. In terms of $\sigma$, this will give the inequality $0\leq\sigma < 2/\epsilon^2-k/2$ which also implies that $0\leq k < 4/\epsilon^2$. Both $k$ and $\sigma$ cannot satisfy their respective upper bounds ($k=4/\epsilon^2$ and $\sigma = 2/\epsilon^2-k/2$) because that will cause the wave function \eqref{WF} to vanish which means the state does not exist (implying that the motion of the system cannot be in any oscillation modes). If $k>4/\epsilon^2$ (which implies $\sigma > 2/\epsilon^2-k/2$) the oscillation frequencies $\Omega_1$ and $\Omega_2$ will become imaginary and the wave function is now not well define. This is the reason that $k \geq 4/\epsilon^2$ and $\sigma \geq 2/\epsilon^2-k/2$ present unphysical situations and have to be excluded from the our consideration. In the physical situations, if we fix the value of the parameter $k$, the entanglement of the system will increase as the parameter $\sigma$ increases and the entanglement will only asymptotically approach $1$, but never reaches $1$, before the parameter $\sigma$ gets to
the cut-off point $\sigma = 2/\epsilon^2-k/2$. Increasing the value of the parameter $k$ will suppress the entanglement of the system like those in the continuous time case.

\section{Concluding discussion}
We can analyse and conclude these results from two different perspectives.

Firstly, if we take the view that the discreteness of time is a fundamental property of the universe, we find that the discrete-time flow affects the system behaviors. Some extra-features, e.g. broader probability contour, modification of the uncertainty principle and cut-off conditions for the ground state entanglement entropy, naturally showed up and will be washed away under the continuum limit \cite{MS}. Interestingly, we find an unexpected relationship between the discrete-time step $\epsilon$, the strength of the mutual interaction between the two subsystems $\sigma$, and the strength of the harmonic potential $k$. In particular, we find that $\sigma$ is bounded from above by a function of $k$ and $k$ is also bounded from above by a function of $\epsilon$. This behavior is completely different from the continuous-time scale ($\epsilon = 0$), where the values of $k$ and $\sigma$ are totally independent.

Secondly, if we look at these results from the operational point of view. By assuming that time is fundamentally continuous but treating that the discreteness arises from experimental samplings of the positions and velocities of the system at a given frequency determined by $1/\epsilon$, we interpret that the difference in the linear entropy for each value of $\epsilon$ is due to the difference in the sampling rate itself. We also discover that the bounds are not actually physical but appear due to the fact that the corresponding cut-off sampling rate ($1/\epsilon_{cut-off} = \sqrt{(\sigma/2)+(k/4)}$) is equal to the Nyquist sampling rate of the system. Thus the reason the observation becomes unphysical beyond that bound is because the sampling rate is less than the Nyquist frequency, which can potentially make the results of the observation become distorted. 

\ack
This work is supported by the Theoretical and Computational Science (TaCS) Center under ComputationaL and Applied Science for Smart Innovation Cluster (CLASSIC), Faculty of Science, KMUTT. WK would like to thank Dr.Thana Sutthibutpong for his help on numerical computation and also to Dr.Ekkarat Pongophas for his helpful discussion. SK is supported by National Research Council of of Thailand (NRCT) under grant No. 219700.

\end{document}